\documentclass[a4paper,11pt]{article}
\usepackage{pos}
\usepackage{graphicx}
\usepackage{wrapfig}
\usepackage{caption}
\usepackage{mwe}
\usepackage{amsmath}
\usepackage{hyperref}
\usepackage{subfig}

\title{Hyperon polarization in deep inelastic scattering}

\author{A. Kerbizi}
\affiliation{Department of Physics, Box 118, 221 00 Lund, Sweden\\
INFN Sezione di Trieste, Via Valerio 2, 34127 Trieste, Italy}

\emailAdd{albi.kerbizi@ts.infn.it}

\def\LambdaBar{\bar{\Lambda}}

\abstract{We have implemented the recent extension of the string+${}^3P_0$ model of spin-dependent hadronization that includes the production of spin-$1/2$ baryons in the Pythia event generator for the simulation of deep inelastic scattering (DIS). The implementation in Pythia is performed by extending the StringSpinner package, which is then used to study the spontaneous polarization of $\Lambda$ and $\LambdaBar$ hyperons produced in the current and target fragmentation regions in DIS with an unpolarized proton target. We find a large spontaneous polarization for $\Lambda$s in the target fragmentation region, calling for the measurement of such observable that can shed light on the quark spin dependence of hadronization.}

\FullConference{
}

\begin{document}
\maketitle

\def\thetacm{\theta^*}
\def\SnT{\textbf{S}_{\rm nT}}
\def\Pn{P_{\rm n}}
\def\PYn{\Pn^{Y}}
\def\n{\textbf{n}}
\def\x{x}
\def\z{z}
\def\PT{P_{\rm T}}
\def\xF{x_F}
\def\GeV{\rm{GeV}}
\def\S{\textbf{S}}
\def\SLambda{\S_{\Lambda}}
\def\SLambda{\S_{\Lambda}}
\def\kT{\textbf{k}_{\rm T}}
\def\zu{\hat{\textbf{z}}}
\def\qu{\hat{\textbf{q}}}
\def\PTvec{\textbf{P}_{\rm T}}
\def\q{\textbf{q}}
\def\Pvec{\textbf{P}_{\Lambda}}
\def\phat{\hat{\textbf{p}}}
\def\Pbf{\textbf{P}}
\def\xF{x_F}
\def\qbar{\bar{q}}
\def\qq{qq}
\def\qqbar{\overline{qq}}
\def\Sq{\textbf{S}_q}
\def\Sqbar{\textbf{S}_{\qbar}}
\def\L{\textbf{L}}
\def\J{\textbf{J}}
\def\Sqq{\textbf{S}_{\qq}}
\def\Sqqbar{\textbf{S}_{\qqbar}}
\def\muqq{\mu_{\qq}}
\def\Im{\rm{Im}}
\def\Ph{\textbf{P}_h}

\def\StringSpinner{\texttt{StringSpinner}}
\def\Pythia{\texttt{Pythia}}
\def\eight{\texttt{8}}

\newcommand\red[1]{{\color{red}#1}}
\newcommand\blue[1]{{\color{blue}#1}}
\newcommand\magenta[1]{{\color{magenta}#1}}
\newcommand\green[1]{{\color{green}#1}}

\section{Introduction}
Large spin-effects are known to exist in the production of $\Lambda$ hyperons since the '70s, when different experiments performing unpolarized hadronic collisions observed $\Lambda$s to be strongly polarized along the vector perpendicular to their production planes. Different phenomenological models were subsequently put forward to explain the puzzling measurements (for a review see, e.g., Ref.~\cite{Panagiotou:1989sv}) . A successful semi-classical model was proposed by the Lund group by using a string model of hadronization and assuming the string breakings to arise via the tunneling of $q\qbar$ pairs in the relative ${}^3P_0$ state~\cite{Andersson:1979wj}. More recently, spontaneously polarized $\Lambda$s were observed in quasi-real photoproduction by the HERMES~\cite{HERMES:2014fmx} and COMPASS~\cite{Grube:2006xia} experiments, and in $e^+e^-$ annihilation by the BELLE experiment~\cite{Belle:2018ttu}. The latter measurement, in particular, led to a renewed interest in the understanding of the origin of the spontaneous polarization of hyperons~\cite{DAlesio:2020wjq,Callos:2020qtu,Kang:2021kpt,DAlesio:2022brl,DAlesio:2023ozw}, which in QCD is explained by the existence of the polarizing fragmentation function (pFF) describing the fragmentation of an unpolarized quark in a transversely polarized baryon~\cite{Mulders:1995dh,Anselmino:2000vs}.

In this work we investigate the not-yet measured spontaneous polarization of $\Lambda$s and $\LambdaBar$s in deep inelastic scattering (DIS) in the context of the string+${}^3P_0$ model of polarized hadronization~\cite{Kerbizi:2021gos}. The latter extends the Lund Model of string fragmentation~\cite{Andersson:1983ia} by including the quark spin degree of freedom at the amplitude level and is implemented in the \Pythia{} Monte Carlo generator~\cite{Bierlich:2022pfr} for the simulation of DIS and $e^+e^-$ annihilation to mesons by the \StringSpinner{} package~\cite{Kerbizi:2023cde,Kerbizi:2026iws}. To enable the simulation of spin effects for hyperon production in DIS we have implemented in \StringSpinner{} the recently extended string+${}^3P_0$ model that includes the production and decays of spin-1/2 baryons alongside mesons~\cite{Kerbizi:2025keh}. The work is organized as follows.
A qualitative description of the string+${}^3P_0$ model with baryon production as well as the implementation in \Pythia{} is given in Sec.~\ref{sec:model}. The results on the spontaneous polarization of $\Lambda$s and $\LambdaBar$s are given in Sec.~\ref{sec:results}. Finally, we conclude in Sec.~\ref{sec:conclusions}.

\section{Inclusion of spin effects for baryon production in \Pythia}\label{sec:model}
\begin{figure}[tbh]
\centering
\subfloat[]{
\begin{minipage}[b]{0.8\textwidth}
\centering
\includegraphics[width=0.65\linewidth]{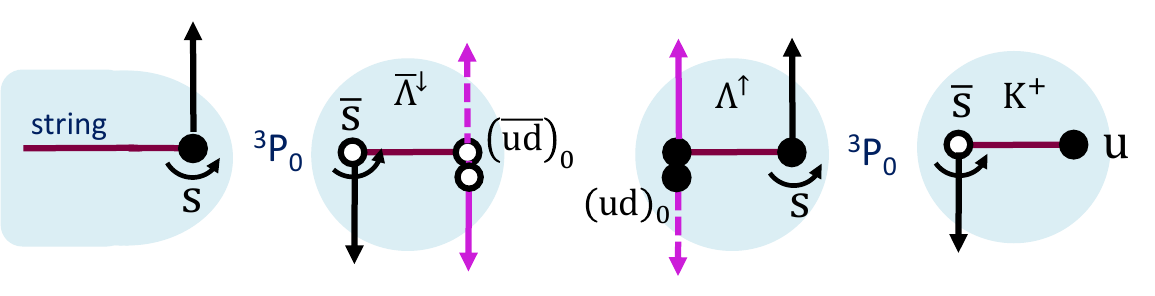}
\end{minipage}
\label{fig:string-CFR}
}
\vspace{-1em}
\begin{minipage}[b]{0.8\textwidth}
  \centering
  \subfloat[]{%
    \includegraphics[width=0.3\linewidth]{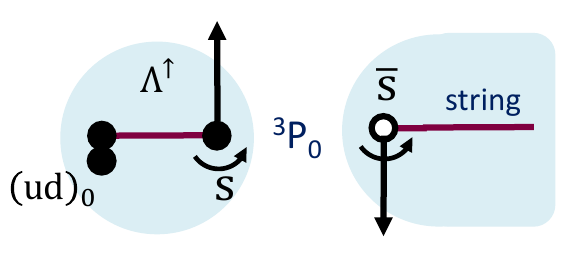}
    \label{fig:string-TFR}
  }
  \hspace{1cm}
  \subfloat[]{%
    \includegraphics[width=0.45\linewidth]{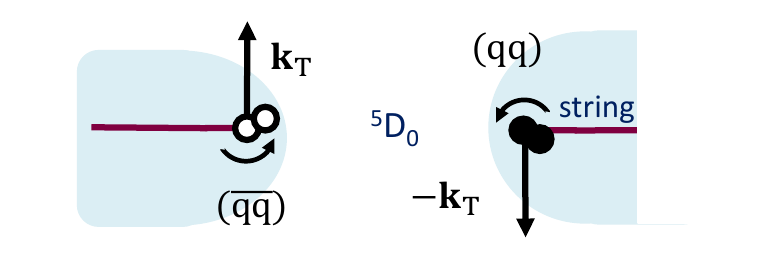}
    \label{fig:5D0}
  }
\end{minipage}
\caption{Fragmentation of the string close to the scattered $u$ quark (a) and to the $(ud)_0$ diquark remnant (b) in  a DIS event with a proton target. Pair production of spin-1 diquarks in the ${}^5D_0$ state (c). The straight (curved) arrows represent the transverse momenta (spins) of quarks or diquarks produced in the string breakings.}
\label{fig:string-CFR-TFR}
\end{figure}

We model the hadronization in a DIS event by the breaking of a relativistic string stretched between the struck quark and the target remnant. For a qualitative description of the extended string+${}^3P_0$ model, let us consider a string stretched by a $u$ quark and a scalar diquark $(ud)_0$, which is a typical configuration for DIS with a proton target. As shown in Fig.~\ref{fig:string-CFR-TFR}, the string can break either by the tunneling of $q\qbar$ pairs or the tunneling of $(\qq)\,(\qqbar)$ pairs. The quarks are produced in the relative ${}^3P_0$ state, namely with unit total spin $\S=\Sq+\Sqbar$ and unit orbital angular momentum $\L$ such that the total angular momentum is $\J=\L+\S=\textbf{0}$. $\J$ is thus locally conserved. The ${}^3P_0$ state induces the correlations $\langle \Sq\cdot(\zu\times\kT)\rangle > 0$ and $\langle \Sqbar\cdot\left[\zu\times(-\kT)\right]\rangle<0$ between the spins $\Sq$ of $q$ and $\Sqbar$ of $\qbar$, and their transverse momenta $\kT$ and $-\kT$ with respect to the string axis. The spin-$\kT$ correlations are responsible for the Collins effect in the string+${}^3P_0$ model~\cite{Kerbizi:2021gos}. In the quantum-mechanical version of the model, which is implemented in the \StringSpinner{} package, the ${}^3P_0$ mechanism is parametrized by the complex parameter $\mu_q$~\cite{Kerbizi:2021gos}~\footnote{The model depends on other two free parameters describing the polarization states of the produced vector mesons.}.


The baryon production in the string+${}^3P_0$ model arises via the tunneling of $(\qq)\,(\qqbar)$ pairs. The diquarks can be either scalar [$(\qq)_0$] or pseudo-vector with spin 1 [($\qq)_1$]. The diquark pair production is suppressed with respect to the quark pair production by $P_{\qq}/P_q\simeq 0.1$~\cite{Bierlich:2022pfr}. Additionally, spin-1 diquarks are further suppressed with respect to scalar diquarks by a factor $P_{(\qq)_1}/P_{(\qq)_0}\simeq 0.03$ on top of a factor of $3$ coming from the counting of spin states~\cite{Bierlich:2022pfr}~\footnote{Note, however, that both parameters $P_{\qq}/P_q$ and $P_{(\qq)_1}/P_{(\qq)_0}$ can be tuned.}. If the string breaks via tunneling of scalar diquarks, the spin of the produced baryon is inherited from the quark coming from the adjacent ${}^3P_0$ breaking. This is the case in Fig.~\ref{fig:string-CFR-TFR} for the production of $\Lambda$s, which is formed by an $s$ quark and a $(ud)_0$ diquark (similarly for $\LambdaBar$).

To conserve locally angular momentum in the tunneling of spin-1 diquarks, the pair is assumed to be produced in the relative ${}^5D_0$ state, namely with total spin $S=2$ and orbital angular momentum $L=2$ such that $\J$ vanishes~\cite{Kerbizi:2025keh}. The ${}^5D_0$ mechanism, shown in Fig.~\ref{fig:5D0}, leads to the correlations $\langle \Sqq\cdot(\zu\times\kT)\rangle > 0$ and $\langle \Sqqbar\cdot\left[\zu\times(-\kT)\right]\rangle<0$ between the spins $\Sqq$ of $(\qq)$ and $\Sqqbar$ of $(\qqbar)$, and the corresponding transverse momenta with respect to the string axis. In the quantum mechanical string+${}^3P_0$ model, the ${}^5D_0$ mechanism is parametrized by a complex parameter $\muqq$ in analogy with $\mu_q$ in the ${}^3P_0$ mechanism. The ${}^5D_0$ mechanism is responsible for a Collins effect in the fragmentation of a transversely polarized quark in a spin-1/2 baryon, which matches the classical picture of the model if $\Im\,\muqq<0$~\cite{Kerbizi:2025keh}.




For the implementation of the spin effects in \Pythia{} we use the rules of the quantum-mechanical string+${}^3P_0$ model in Ref.~\cite{Kerbizi:2025keh}. In particular the hadronization chain is simulated by recursive $Q\rightarrow h+Q'$ splittings, where the fragmenting particle is $Q=q,(\qq)$, the emitted hadron is either a meson $M$ or a spin-1/2 baryon $B$, and the leftover particle is $Q'=q'$ if $h=M$ or $Q'=\qqbar$ if $h=B$~\footnote{The charge conjugated splittings are also taken into account.}. Each splitting is described in momentum and spin space by a splitting matrix $T_{Q',h,Q}$, which we have used to implement the model in \StringSpinner~\footnote{The splitting matrix allows for the calculation of the probability for the polarized splitting to occur and to propagate the spin effects along the hadronization chain, as done in Refs.~\cite{Kerbizi:2023cde,Kerbizi:2026iws} for the production of mesons.}.
Concerning the free parameters of the string+${}^3P_0$ model we use the setting in Ref.~\cite{Kerbizi:2024vpd} and take $\muqq = \mu_q^*$.

The detailed description of the implementation in \StringSpinner{} will be given in a separate work.

\section{Results on the spontaneous polarization}\label{sec:results}
\subsection{The extraction method}
\begin{figure}[h]
\centering
\includegraphics[width=0.45\textwidth]{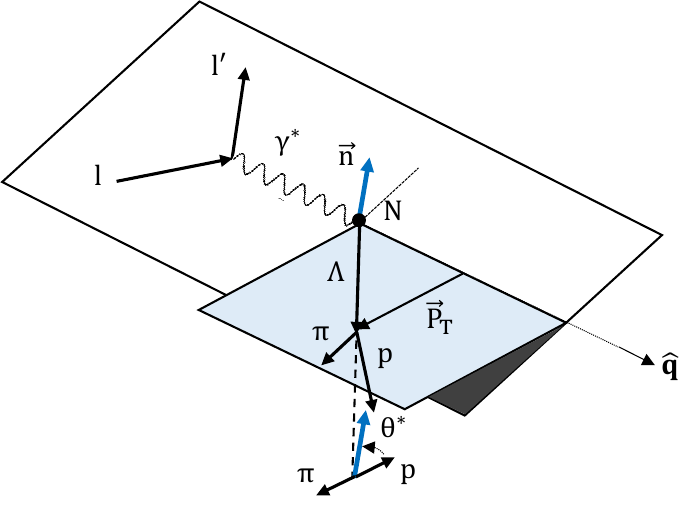}
\caption{The kinematics of the SIDIS process in the GNS for the production and decay of a $\Lambda$ hyperon.}
\label{fig:SIDIS}
\end{figure}

We consider the production of a hyperon $Y=\Lambda,\LambdaBar$ in DIS in the so-called gamma-nucleon reference system (GNS), shown in Fig.~\ref{fig:SIDIS}. The momentum $\q$ of the exchanged virtual photon defines the longitudinal direction. The transverse momentum of the hyperon with respect to $\q$ is indicated by $\PTvec$. The vector perpendicular to the production plane of the hyperon, spanned by $q$ and $\PTvec$, is given by $\n=(\q\times \PTvec)/|\q\times \PTvec|$.


For $Y=\Lambda,\LambdaBar$, the spontaneous polarization $\PYn$ can be analyzed by the weak decays $\Lambda^{\uparrow}\rightarrow p+\pi^-$ and $\LambdaBar^{\uparrow}\rightarrow \bar{p}+\pi^+$. The angular distribution of the decay baryon $B=p,\bar{p}$ in the rest frame of $Y$, reached from GNS by a boost along the hyperons momentum, is
\begin{equation}\label{eq:N}
    N_p^Y(\cos\thetacm;X)\propto 1+\alpha_{Y}\,\PYn\,\cos\thetacm.
\end{equation}
$\thetacm$ is the angle between the momentum of $B$ and the vector $\n$ perpendicular to the hyperons production plane. $\alpha_Y$ is the decay constant parameterizing the weak decay $Y$. The considered kinematic variables are $X=\x,\z,\PT,\xF$, where $\x$ is the Bjorken variable while $z$ and $\xF$ are respectively the fractional energy and the Feynman-$x$ variable of the hyperon.


For a given hyperon $Y$ and an interval for the kinematic variable $X$, the distribution of the decay baryon $B$ is constructed and it is fitted by the function $f(\cos\thetacm)=p_0\times(1+p_1\,\cos\thetacm)$ motivated by Eq.~(\ref{eq:N}). The spontaneous polarization is finally evaluated as $\PYn=\hat{p}_1/\alpha_Y$, where $\hat{p}_1$ is the value of the parameter resulting from the fit.

In this work we consider the kinematic configuration of the COMPASS experiment, namely a $160\,\GeV$ muon beam and unpolarized proton target at rest~\cite{COMPASS:2014bze}. Inspired by the analysis of transverse single-spin asymmetries in Ref.~\cite{COMPASS:2014bze}, we require the virtuality of the exchanged photon to be $Q^2>1\,(\GeV/c)^2$, the mass of the final state hadronic system to be $W>5\,\GeV/c^2$, and the inelasticity to be $0.2<y<0.9$.

\subsection{Polarization of $\Lambda$ and $\LambdaBar$ hyperons}
\begin{figure}[t]
\centering
\begin{minipage}[b]{0.75\textwidth}
\includegraphics[width=1.0\textwidth]{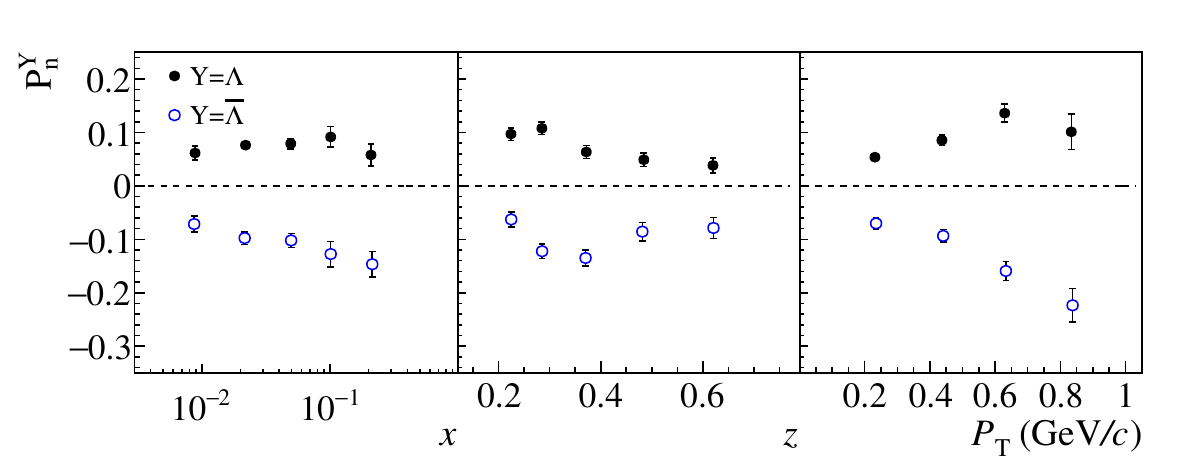}
\end{minipage}
\caption{Simulated spontaneous polarization of $\Lambda$ and $\LambdaBar$ hyperons as a function of $\x$, $\z$ and $\PT$.}
\label{fig:xzPT}
\end{figure}

Figure~\ref{fig:xzPT} shows the spontaneous polarization of $\Lambda$ (full circles) and $\LambdaBar$ (open circles) hyperons extracted as a function of $\x$, $\z$ and $\PT$. We have applied the kinematic selections $z>0.2$ and $\PT>0.1\,\GeV/c$. 

As can be seen, the polarization of $\Lambda$ hyperons has positive sign whereas the polarization of $\LambdaBar$ hyperons has negative sign. It reaches an absolute value of above $10\%$. The sign of the polarization can be understood from Fig.~\ref{fig:string-CFR}. Due to the suppression of spin-1 diquarks, polarized $\Lambda$s are most likely produced after a kaon. They are composed of an $s$ quark, which carries the spin of the hyperon, and a scalar $(ud)_0$ diquark. The spin of the $s$ quark is correlated with the transverse momentum $\kT$ of the quark due to the ${}^3P_0$ mechanism. This leads to $\langle \S_{\Lambda}\cdot (\zu\times \PTvec)\rangle>0$, namely to a positive spontaneous polarization for the $\Lambda$. Leading unpolarized $\Lambda$s are formed by the scattered $u$ quark and a $(ds)_0$ diquark produced in the string breaking. They dilute the spontaneous polarization at large $z$ in Fig.~\ref{fig:xzPT}. Similar considerations lead to $\langle \S_{\LambdaBar}\cdot (\zu\times \PTvec)\rangle<0$ and therefore to a negative spontaneous polarization for the $\LambdaBar$. 

$\Lambda$s and $\LambdaBar$s can also be produced in the hyperon decays, e.g. $\Sigma^0\rightarrow \Lambda+\gamma$ and $\bar{\Sigma}^0\rightarrow \bar{\Lambda}+\gamma$, which dilute the measured spontaneous polarization. We find that about half of $\Lambda$s and half of $\LambdaBar$s come from decays of heavier hyperons. 



Note that the $\Lambda$ and $\LambdaBar$ polarizations as a function of $\z$ in Fig.~\ref{fig:xzPT} are similar to the predictions performed by different groups for this observable in the kinematic configuration of the Electron Ion Collider~\cite{Kang:2021kpt,DAlesio:2023ozw}.


\begin{figure}[h]
\centering
\begin{minipage}[b]{0.55\textwidth}
\includegraphics[width=1.0\textwidth]{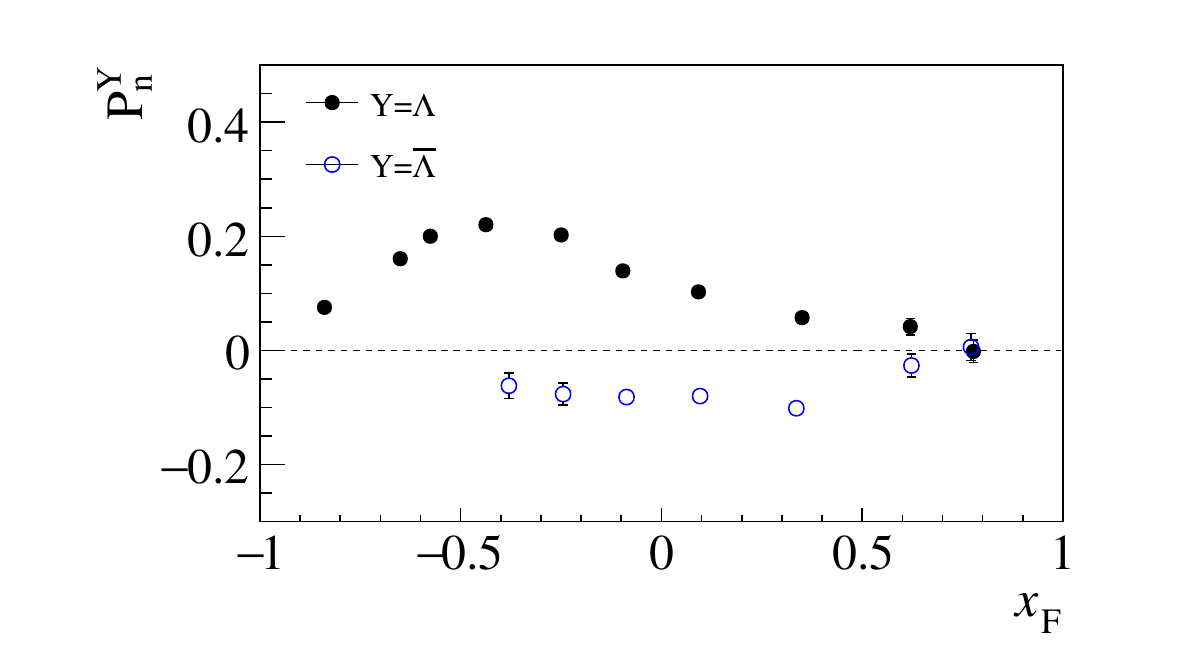}
\end{minipage}
\caption{Simulated spontaneous polarization of $\Lambda$ and $\LambdaBar$ hyperons as a function of $\xF$.}
\label{fig:xF}
\end{figure}

In Fig.~\ref{fig:xF} we show the spontaneous polarization of $\Lambda$ (closed circles) and $\LambdaBar$ (open circles) as a function of $x_F$. In order to study the polarization close to the target-remnant fragmentation region taking place at negative $x_F$ values, we have removed the selection $z>0.2$. As can be seen, the $\Lambda$ polarization has a striking strong dependence on $\xF$: it increases up to about $20\%$ for $\xF\sim -0.5$ and decreases for lower $\xF$~\footnote{This result suggests a large fracture function $\Delta_{\rm{T}}\hat{M}^h$~\cite{Anselmino:2011ss}.}. This is due to the combination of two main effects. Since a $u$ quark is most likely struck in the DIS event, the remainder of the proton target is mainly $(ud)_0$. This makes possible the production of a $\Lambda$ composed by the $(ud)_0$ diquark and an $s$ quark from $s\bar{s}$ pair produced in the ${}^3P_0$ state, as shown in Fig.~\ref{fig:string-TFR} (see also Ref.~\cite{Andersson:1979wj}). This leads to a $\Lambda$ with positive polarization. In another likely event for $\Lambda$ production a $\bar{s}$ quark is instead struck. The target remnant is a four-quark state $|suud\rangle$ that is split by Pythia in a baryon $B$ and a remainder quark without the spin effects. The case $B=\Lambda$ leads to the dilution of the spontaneous polarization of $\Lambda$s for $\xF\lesssim -0.5$. $\LambdaBar$ hyperons, instead, are not produced among the target fragments and the increase of the spontaneous polarization for negative $\xF$ is not observed, as seen in Fig.~\ref{fig:xF}.

Simulations of DIS events in the kinematics of the HERMES experiment lead to similar results as in Fig.~\ref{fig:xzPT} and in Fig.~\ref{fig:xF}.


\section{Conclusions}\label{sec:conclusions}
We have studied the spontaneous polarization of $\Lambda$ and $\LambdaBar$ hyperons in DIS events with an unpolarized proton target by interfacing the string+${}^3P_0$ model in Ref.~\cite{Kerbizi:2025keh} to the \Pythia{} generator. This is achieved by an extension of the \StringSpinner{} package. The model predicts large spontaneous polarizations for $\Lambda$ and $\LambdaBar$ hyperons and a strong dependence on $z$, $\PT$ and $\xF$. An interestingly large spontaneous polarization is predicted for $\Lambda$s with $\xF<0$, namely near the target remnants. The obtained results call for a measurement of the hyperon spontaneous polarization in DIS, which could shed new light on the spin effects in hadronization.

\subsection*{Acknowledgments}
I would like to thank Leif Lönnblad and Gösta Gustafson for the valuable discussions and advise on this work. The work was done in the context of the project “SPINFRAG: Spin-dependent string fragmentation”, funded by the European Union under Marie Skłodowska-Curie Actions (MSCA), grant agreement ID 101107452.

\end{document}